\documentclass[10pt,a4paper]{article}
\usepackage{amsmath}
\usepackage{amsthm,amssymb}

\numberwithin{equation}{section}
\usepackage{graphicx}

\title{\Large \bf  Intrinsic Classes in the  Union of European Football Associations Soccer Team Ranking  }

\author{ \large \bf  Marcel    Ausloos$^{1,*,2,}$
 \\ \\$^{1,*}$  Royal Netherlands Academy of Arts and Sciences\footnote{Associate researcher}, \\Joan Muyskenweg 25, 1096 CJ Amsterdam, The Netherlands \\  
\\ $^2$ GRAPES, rue de la Belle Jardini\`ere 483, B-4031 Li\`ege, \\Federation Wallonie-Bruxelles, Belgium  \\  email: marcel.ausloos@ulg.ac.be\\
}

\begin{document}
\maketitle
\begin{abstract}
A   strong structural   regularity of  classes is found in soccer teams ranked by  the  Union of European Football Associations (UEFA) for the  time interval 2009-2014. It concerns 424 to 453 teams according to the  5 competition seasons.   The analysis is based on the rank-size theory considerations, the size being the UEFA coefficient at the end  of a season.    Three classes emerge: (i) the few "top" teams, (ii)  300 teams, (iii) the rest of the involved teams (about 150) in the tail of the distribution.  
There are marked empirical  laws describing each class. A 3-parameter Lavalette function is used to describe the concave curving as the rank   increases, and to distinguish the the tail   from the central behavior.     
\end{abstract}

 keywords: team ranking \*\ soccer \*\ rank-size relation \*\ Lavalette function \*\ intrinsic complexity

\section{Introduction}\label{Introduction}
\par
Nonlinearity   and complexity  are common features of a  large number of systems studied in  modern science  \cite{compl1}-\cite{roehner2007driving}.  In many cases,  researchers have detected the existence of power laws,  for different characteristic quantities of  such complex systems.    
These interesting  contributions, at the interfaces of various disciplines,  are often tied to various technical questions or are limited to the analysis of distribution functions,   themselves considered to be the first to look at for characteristics of  complex systems, but without conveying questions tied to self-organizations \cite{soc} or external constraints \cite{roehner2007driving}. 

In particular, ranking analysis has   received much attention,  
   since  Zipf \cite{z1}  observed  that a large number 
of  size  distributions, $N_r$ can  be approximated by a simple  {\it 
scaling (power) law} $N_r = N_1/r $,   where  $r$ is the ranking parameter, with 
 $N_{r } \ge N_{r+1}$, (and obviously  $r<r+1$). 
   Zipf\'s idea has led to a flurry of log-log diagrams showing a straight line through the displayed data  such that   "any" size distribution,   more generally  called $y_r $,  reads
  \begin{equation}\label{Zipfeq}
y_r = \cfrac{a}{r^\alpha} \;,
\end{equation}
i.e., the so called rank-size scaling law. 
 The scaling exponent  $\alpha$ is considered to indicate whether  the size distribution $y_r $ is close or not  to  some optimum  (equilibrium) state \cite{z1}, i.e. when $\alpha=1$.  The amplitude $a$ can be estimated from the normalization condition. Indeed, 
the pure power-law distribution, for a $continuous$ variable,  known as the $zeta$-distribution  \cite{Edwards,Temme,Titchmarsh},   or discrete Pareto distribution, reads
\begin{equation}\label{eq0} p(k) = \frac{k^{-\gamma}}{\zeta(\gamma)}
   \end{equation} 
where    $p(k)$ is the probability of observing the value $k$, a positive integer, 
  $\gamma$  is the power-law exponent,  and $\zeta(\gamma)$ $\equiv$$ \sum_{k=1}^{\infty} k^{-\gamma} $ is the Riemann zeta function; note that $\gamma$, in Eq.(\ref{eq0}) must be greater than 1 for the Riemann $zeta$-function to be finite. Therefore, for the discrete distribution,  Eq.(\ref{Zipfeq}), $a\simeq k_M/ \zeta(\alpha) \sim k_M/2$, where $k_M$ is the largest value of $k$.

This might be the case in sport competition ranking, though the number of scales is obviously finite.     Here below, an analysis of  data from a specific nonlinear complex system, i.e.  the  Union of European Football Associations (UEFA) team ranking,  as a specific modern society interesting example, is  discussed.


 The data is described and its  analysis performed  in Sect. \ref{UEFAdata} through  simple empirical laws  in order to introduce possible fits.
The classical  rank-size, hyperbolic, Eq.(\ref{Zipfeq}), relationship is  not found  for  UEFA teams. On the contrary, after  many  data statistical tests  (not reported), classes emerge, best seen through the use of a 3-parameter, thus generalized, Lavalette function.  A top, a middle and a low class  of teams appear.
 Several remarks  serve as conclusions, in Sect. \ref{conclusions}.

Warning:  it should be obvious that the ranking   of a team may change  from a   year to another, after some season; see Appendix. There is $no$ further consideration here on the time  evolution of a team through the ranking. No doubt that the time dependence of the ranks  should be of interest as well, but, due to likely economic conditions, beside  sport ones, such a subject  is left for  dynamic evolution studies, including evolution modeling, outside the present aims.

Note, in concluding this introduction, that the literature on  $sport$ ranking is  very large, in particular  tied to economics questions, as in \cite{NJP14.12093038scoreprize,pilavci2011pitch,barros2008identification,barros2008efficiency}. Thus, we mention  a few publications where soccer and ranking from direct measures (win, draw, loss) have been considered:
\begin{itemize}
 \item
 Stefani  1997 pioneering survey of the major world sports rating systems, including  soccer through FIFA\footnote{ At the time of writing, FIFA  \cite{FIFA}, is made of  6 confederations \cite{Confacronym}, grouping  209 Member Associations squads,  $\sim"countries"$,   53 of them being in  the UEFA.}  rules,  is first  to be noted \cite{JAS24.97.635worldsportsratingStefani};
  \item  Kern and   Paulusma 2001  paper  discussing FIFA rules  complexity  for   competition outcomes, leading to  $team$ ranking  \cite{DAM108.01.317}, from where
  \item Macmillan and  Smith, explaining $country$  ranking in 2007  \cite{JSE08.07.202explainingrankingsoccercountries}; such a theme being reconsidered by
  \item  Ausloos et al. \cite{IJMPCFIFAMARCAGNV}   comparing the FIFA  country ranking, based on games between national  squads,  to the \underline{country} UEFA ranking, based on $team$ game results, and 
  \item  Constantinou and Fenton \cite{pi-ratings}    determining  the level of "ability" of  (five English Premier League seasons) soccer teams  based on the relative discrepancies in scores between adversaries. 
    \end{itemize}

\section{UEFA team  coefficient  data }\label{UEFAdata}

Usually, the ranking represents the overall performance over the period of  5 consecutive seasons \cite{UEFAteamrankingrules}, after averaging points obtained in various competitions, at the end of each  year (more exactly season), only for teams having participated in the  UEFA Champions League and  the UEFA Europa League. Thus, e.g.   the 2009/10  coefficient results from games having taken place in 2005/06 $\dots$, 2009/10. The rules  are more complicated than  a "win-draw-loss"  rating.  They depend on the success at some competition level, and differ according to the competition.  A UEFA coefficient  table is freely available and is updated regularly depending on the competition timing.  
 Here below,  5 different consecutive seasons data are examined: 2009/10, $\dots$, 2013/14 (till May 2014).

 The number of concerned  teams  ranges from 424 to   453 according to the season. In statistical physics terms, this is an open system, with birth and death processes. However such events mainly occur for the "not too top" teams.
 The statistical characteristics of the  ranking distributions for the various years are given in Table  \ref{Table0914stat}.  It can be noticed that the mean is  increasing rather  slowly,  as do the skewness and kurtosis, indicating a widening of the distributions, and a kind of moving average effect, but   $\mu/\sigma\sim 0.64$ is rather stable indicating some shuffling mainly among the top teams.
    \begin{table} \begin{center} 
\begin{tabular}[t]{cccccccccc} 
\hline 
 UEFA coeff. &	 09/10 &10/11 &11/12 &12/13 &13/14\\\hline
N. of teams	 & 	424	 & 	439	 & 	443	 & 	450	 & 	453\\
Minimum	 & 	0.150  & 	0.183	 & 	0.183	 & 	0.133	 & 	0.449\\
Maximum	 & 	136.951	 & 	151.157	 & 	157.83701	 & 	157.605	 & 	159.456\\
Mean	 ($\mu $) & 	15.293 & 	15.359  & 	15.693 & 	16.050 & 	16.662 \\
Median	 & 	4.838 & 	4.825& 	5.180& 	5.809	 & 	6.825\\
RMS	 & 	27.95 & 	28.67 & 	29.35 	&29.75 & 	30.54 \\
Var  ($ \sigma^2$) & 	548.85 & 	587.15 & 	616.47& 	628.91 & 	656.61 \\
Std Error	 & 	1.138& 	1.156 & 	1.180& 	1.182 & 	1.2047\\
Skewness	 & 	2.519 & 	2.669& 	2.716 & 	2.745	 & 	2.917 \\
Kurtosis	 & 	6.799& 	7.712& 	8.072  & 	8.337 & 	9.621\\\hline

 $\mu/\sigma$ &0.653&0.634&0.632 &0.640 &0.650 \\
  \hline
\end{tabular} 
   \caption{Summary of  statistical  characteristics  for UEFA team ranking  coefficient data   }\label{Table0914stat}
\end{center} \end{table}

 \subsection{Empirical Ranking Laws}\label{rankinglaws}
Beside the classical  2-parameter power law, Eq.(\ref{Zipfeq}), 
  \begin{itemize}
  \item the mere exponential  (2 parameter fit ($b$, $\beta$) case  
  \begin{equation} \label{expL}
 y(r)= b \;e^{-\beta r}
\end{equation}
 and  the  Lavalette 2-parameter free  ($\kappa_2$, $ \gamma$) form, when  the data crushes at high $x$-axis value,  as it results from a finite size of the  number  $N$ of system elements 
\cite{Lavalette},
\begin{equation} \label{Lavalette2} 
y(r)= \kappa_2\; \big[\frac{N\;r}{ N-r+1} \big] ^{- \gamma}  
\end{equation} 
\end{itemize}
 and  3-parameter  statistical distributions, like
  \begin{itemize}
\item
  the power law with cut-off  \cite{Pwco3}:
  \begin{equation} \label{PWLwithcutoff}
 y(r)= c \;r^{-\lambda} \; e^{-\zeta r}.
\end{equation}
\item  and a mere generalization of Eq. (\ref{Lavalette2}), i.e.,
 allowing  for two  different exponents ($ \gamma$ and $\xi $) at low and high ranks \cite{JoI1.07.155Mansilla,JQL18.11.274Voloshynovska}:
  \begin{equation} \label{Lavalette3a}
 \;\;  y_N(r)= \kappa_3\;  \frac{(N\;r)^{- \gamma}}  { (N-r+1)^{-\xi}  }
\end{equation}
 
\end{itemize}
 should be also  considered.

Note that, in Eq. (\ref{Lavalette2}) and Eq. (\ref{Lavalette3a}),  the role of $r$ as the independent variable, in Eq.(\ref{Zipfeq}), is taken by the ratio $r/(N - r + 1)$ between the descending and the ascending ranking numbers. Moreover, practically at data fitting time,  one can also use
 \begin{equation} \label{Lavalette3u}
\;\;  y(u)
\equiv  \hat{\Lambda}\;   u^{-\phi}\;(1-u)^{+\psi}\;
\end{equation}
with $u=r/(N+1)$, emphasizing a sort of universality form. For   $\psi=0$, it   reduces to Eq.(\ref{Zipfeq}). Observe that  the slope on a log-log plot in the central region, at $u=1/2$,  is equal to $-2(\phi+\psi)$.

    \begin{table} \begin{center} 
\begin{tabular}[t]{cccccccccc} 
\hline 
  &	 09/10 &10/11 &11/12 &12/13 &13/14\\\hline
$N$ ($\equiv$ d+1) & 	424	 & 	439	 & 	443	 & 	450	 & 	453\\
$\hat{\Lambda}$ & 45.58&47.34 & 48.58 & 47.60& 43.29\\
$\phi$& 	 0.201 &  0.205 &  0.206&  0.214 & 0.241 \\
$\psi$& 4.053& 4.455 & 4.557 &4.326& 4.104\\
$\chi^2$ &1878.5&2018.4&2378.0& 3041.0 &4.531.8\\
$R^2$ & 0.992& 0.992&0.991&0.989 & 0.985 \\
 \hline
$ \alpha$&0.11&0.20&0.17&0.11&0.05\\
$N$ top&7&7&6&5&3\\\hline
$\phi$& 0.37&0.37    &  0.30 &0.31    &0.37   \\
$\psi$& 2.83 &3.16   & 3.68 &3.51 &2.84   \\\hline
\end{tabular} 
   \caption{Summary of  parameter values in Lavalette 3-parameter free reduced form, Eq. (\ref{Lavalette3u}), for UEFA team ranking data \underline{overall} $N$ ranges  in each season;  d is the number of degrees of freedom (=$N-1$) in a $\chi^2$ test; the data corresponds to Fig. \ref{screenshotPlot1UEFA5yrs5u}. The $\alpha$ exponent results from a power law fit to the ranked top team coefficients. The last 2 lines refer to  a 3-parameter Lavalette function  fit to the middle class and low class team data   }\label{Table1UEFA5yrs5u}
\end{center} \end{table}

  \subsection{Data Analysis}\label{dataanalysis}

\begin{figure}
 \includegraphics [height=13.0cm,width=13.0cm]{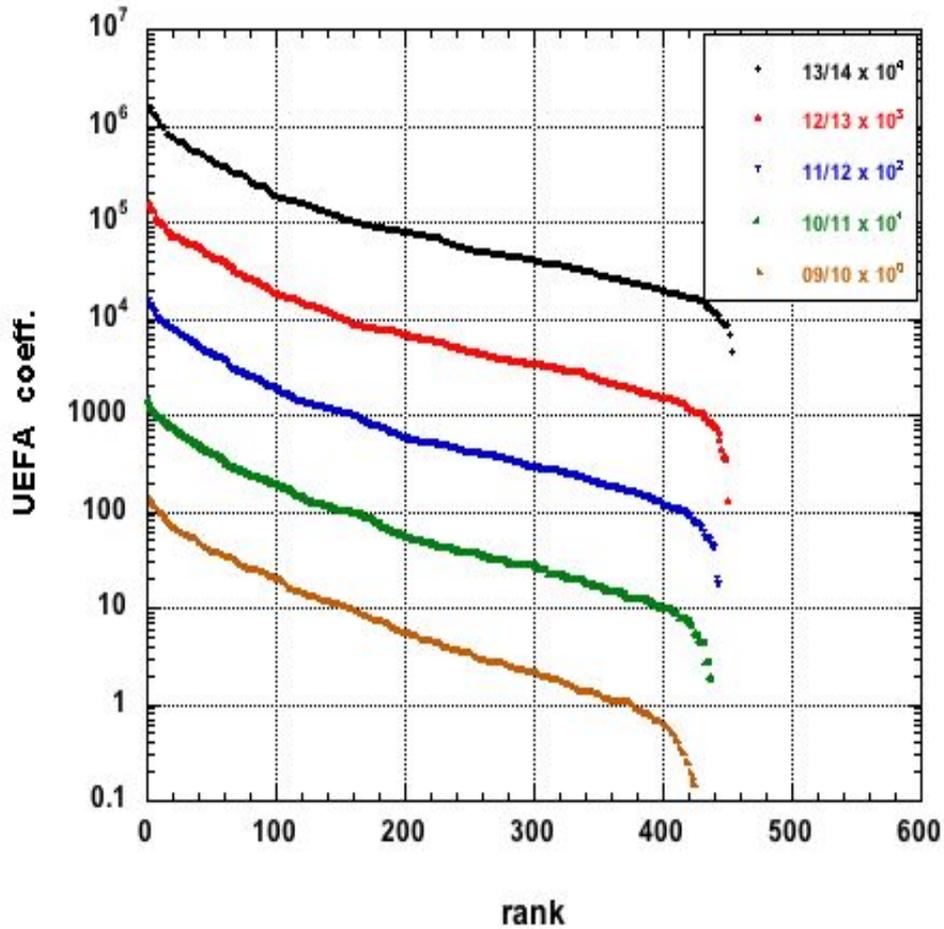}
 \caption   {Yearly ranking of the teams as a function of their UEFA coefficient:  different colors and symbols are  used for  different seasons; the coefficient values have been multiplied as indicated in the inset to make the data distinguishable }    \label{Plot14teamrank0914lilo}
 \end{figure}

The yearly ranking of the teams as a function of their UEFA coefficient is shown in Fig. \ref{Plot14teamrank0914lilo}, on a \underline{semi-log plot};  a  different color and symbol  are   used for each year; the coefficient values have been multiplied as indicated in the inset to make the data  readable. The change in curvature (near $r= 200$) suggests consideration of  a  Lavalette function for describing the data rather than a mere power law or exponential law, or their  product as in Eq.(\ref{PWLwithcutoff}).  These 3 laws would imply a tail at high rank. 

Nevertheless, a sharp change  in behavior can be  noticed  at  very low ranks when a smooth line is drawn through the data at first. Indeed, the derivative of this guiding line for the eye has a sharp peak at a $r_1$ value given in Table 2. Below this rank value, the best empirical law fitting the data is markedly a power law, as found on a log-log plot (not shown). The exponent is given in Table 2 as well.

 For completeness, note that  the Levenberg-Marquardt algorithm  \cite{Levenberg44,Marquardt63,LMalgorithm,ranganathan2004levenberg} has been used  for the fitting procedure of the data to the mentioned non-linear functions.  
 
 \subsection{Rank Classes}\label{discussion}

\begin{figure}
 \includegraphics [height=13.0cm,width=13.0cm]{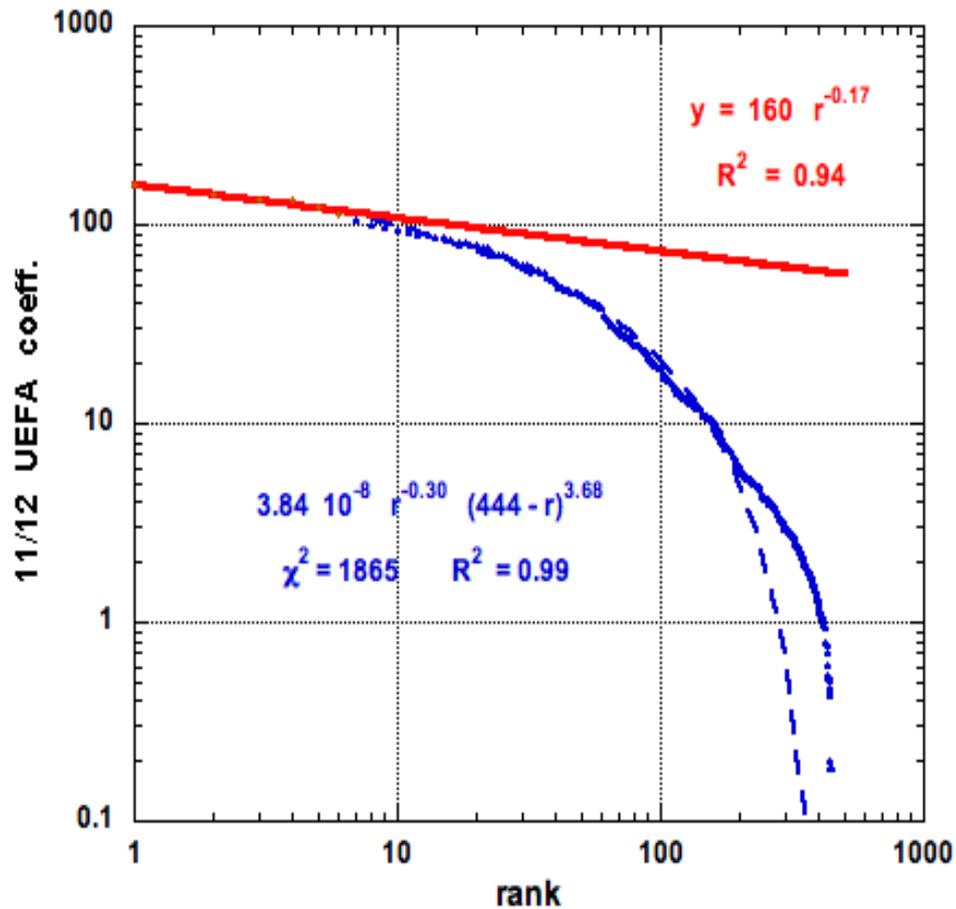}
 \caption   { 
  Log-log plot of the 2011/12 ranking  distribution of the teams  due to  their UEFA coefficient:  the straight  (red) line is a power law fits to the 6 top teams; it is followed by a 3-parameter Lavalette (blue dash lines)  function fit to the other 436 teams, a fit which deviates from the data at a shoulder rank $r\sim$  170. Nevertheless, $\chi^2$ and $R^2$ values indicate  very significant fits      }    \label{ScreenshotPlot24fig2retried}
 \end{figure}

 Next, consider  an example, as in   
Fig.  \ref{ScreenshotPlot24fig2retried}, i.e. the
 2011/12 ranking  distribution of the teams  due to  their UEFA coefficient, on a   \underline{log-log plot}. An overall fit by a generalized  Lavalette function implies a too strong importance of the low rank coefficients, curving the fit line too strongly at moderate ranks (not shown).  Observing  the relatively well pronounced hyperbolic shape at very low rank,   a straight  (red) line,  i.e.,  a power law fits can be made for the 6 top teams. Thereafter removing such "top  class" teams from the fit, a Lavalette function fit can be attempted, as indicated by the blue dash lines. Note that the  fit   deviates  again from the data at a shoulder rank $r\sim$  170. Nevertheless, the $\chi^2$ and $R^2$ values indicate  very significant fits. Similar considerations hold for the other seasons, but the data is not displayed for conciseness. It is available from the author  upon request.  A point should be  emphasized here. Observe the change in the power law exponent between the low ranks and the medium rank ranges:  from 0.17  (except for 2013/14, - which might be due to incomplete data at the time of downloading) to 0.37.  Approximately the same  values and changes occur  in  the other 3 cases. Note the high value of the $\psi$ exponent, i.e. $\sim 3.0$.

\begin{figure}
 \includegraphics [height=13.0cm,width=13.0cm]{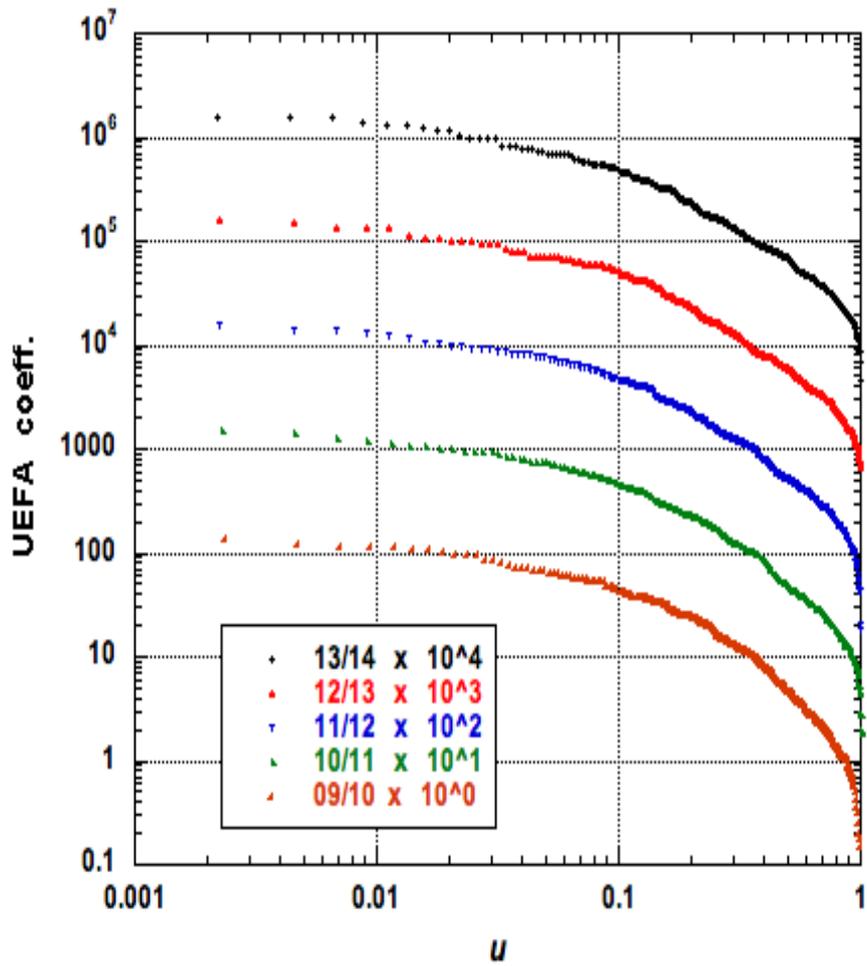} 
 \caption   { 
  Log-log plot of the  ranking  distribution of the teams  due to their UEFA coefficient   as the function of the  universal  variable $u$:  different colors and symbols are  used for  different seasons; the coefficient values have been multiplied as indicated in the inset to make the data distinguishable }    \label{ScreenshotPlot4UEFA5yrsshftdu} 
 \end{figure}

   In order to indicate 
the "universality" of the findings, let the $u$ variable, introduced here above be considered for the ranking.
A log-log plot of the  ranking  distributions of the teams  due to their UEFA coefficient   as the function of the  universal  variable $u$ is shown in   Fig.\ref{ScreenshotPlot4UEFA5yrsshftdu}:  different colors and symbols are  used for  different seasons; the coefficient values have been multiplied as indicated in the inset to make the data distinguishable. The consistency is remarkable. The figure allows to emphasize the different regimes, at $\sim$ 0.015 and $\sim$ 0.20 in all cases. There are finer structures, but not so obvious ones, likely due to some reshuffling effects of the team ranks, and the sort of moving average which occurs when calculating the season rank from the five season coefficients.

\begin{figure}
 \includegraphics [height=13.0cm,width=13.0cm]{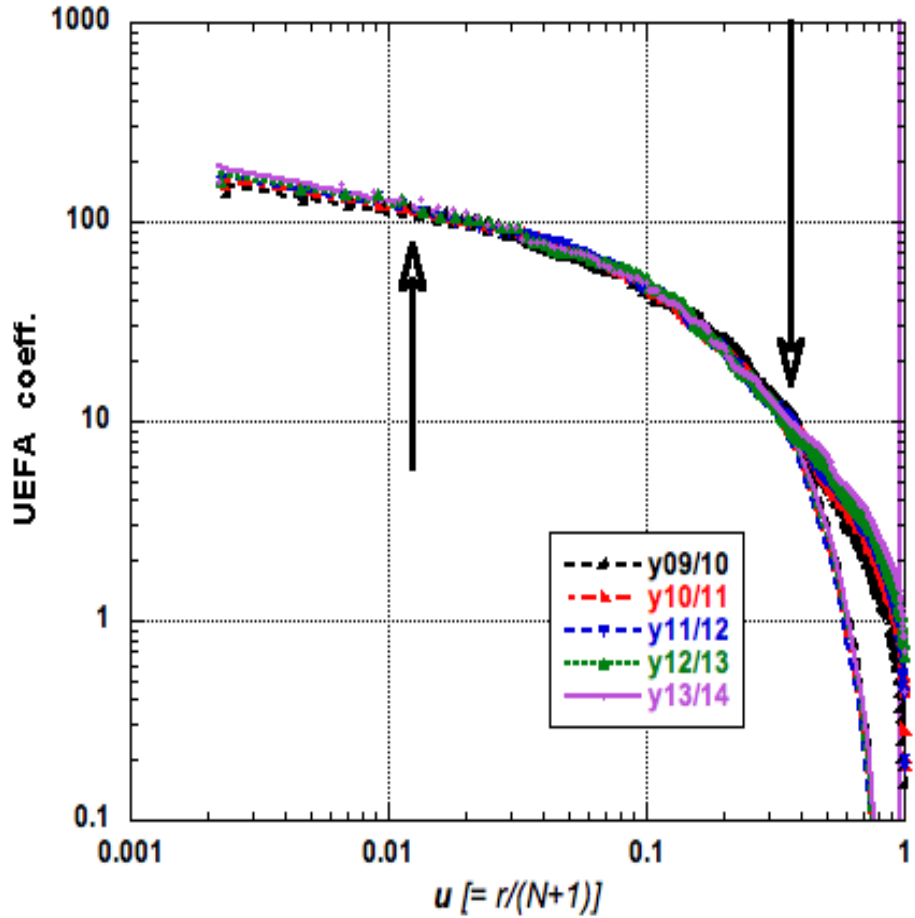} 
 \caption   { Universal
  log-log plot of the  ranking  distribution of the teams due to their UEFA coefficient;  different colors and symbols are  used for  different seasons (ynn/mm);    the straight   power law fits for  the  top teams  are not emphasized; the 3-parameter Lavalette (colored  dot or dash lines)  functions  are  shown; arrows indicate deviations between fits and data, defining  the  low rank  ($u\sim 0.012$, i.e.   $r\simeq  6 $  )  regimes, the middle class regime, and  the low class regime after a shoulder  for $u\sim 0.35$, i.e.   $r\simeq 160 $  }    \label{screenshotPlot1UEFA5yrs5u} 
 \end{figure}

   The final "proof" of the  classes is found in Fig.   \ref{screenshotPlot1UEFA5yrs5u} through a
   log-log  display of the  ranking  distribution of the teams due to their UEFA coefficient. The "universality" is convincing. The  arrows indicate deviations between fits and data, defining  the  low rank  ($u\sim 0.012$, i.e.   $r\simeq  6 $  )  "top class" teams, the "middle class" team regime, and  the "low class"  team regime after a shoulder  for $u\sim 0.35$, i.e.   $r\simeq 160 $.

\section{Concluding remarks}\label{conclusions}

The classical rank-size relationship \cite{newman,Brakman}, underlying  the description of social   complex    systems, has been examined  for the ranking of  soccer teams in UEFA competitions. The  indicator  of  such a {\it  sport  team class system} has been considered to be   the team UEFA coefficient. The UEFA coefficients originate from complicated rules (not discussed here) which seem to imply the creation of team classes. A  sharp conclusion  is first reached that the distribution of  UEFA team ranking does $not$ follow   a single power law, nor an exponential, in fact. Instead of this, it appears that the   rank-size  distribution contains 3  classes, approximated by a mere  scaling (power) law,  with a  quite different exponent,  $\simeq$ 1/5 or 1/3 - suggesting a sort of order-disorder phase transition, in a thermodynamics-like sense, between the low and  medium rank teams.  Moreover, through this log-log search for an empirical law, attempting a Lavalette function, as in other informetrics systems,  the middle class is enhanced. As a moral conclusion, it seems that the UEFA rules are close to favorize a sort of Matthew effect, for  the top (7 or so) teams, - which is not without recalling economic considerations \cite{NJP14.12093038scoreprize,pilavci2011pitch,barros2008identification,barros2008efficiency}. 
 
It is commonly accepted that the rich teams  are  those which are better ranked. There is no study here about the correlation between  team richness nd UEFA ranking.  It is often  considered that there is some correlation, but this not completely proven \cite{deloittetouche2013}. Indeed,   the first top 10 teams
do not fall permanently  in the top class, as shown in the Appendix.  Note that Paris St Germain is  a typical outlier in this respect, not appearing in the top 10.  If  the content of the  top class changes from one season to another,  it is mainly due to the "organizing rules" for UEFA ranking, based on 5 years results in specific competitions. Since the ranking is much due to Champion's and Europa League games,  indeed the  ranking stems from results in  such   competitions.

In this respect, some physics modeling can be suggested for future work, along the lines of open systems in a necessarily   non-equilibrium state. Recall that the UEFA ranking allows new teams "to come in" every year. A  few can move out after 5 years. By considering a non-equilibrium ensemble of many replicas as a large, and therefore thermodynamic, system,
extensive variables  can be defined  such that   well-known
thermodynamic concepts and relations can be applied to small
systems.   Rubi and coworkers,    following Hill's line of thought   \cite{Hillbook1994,adamrubi}  have shown
   how to construct  such a nonequilibrium thermodynamics for  finite size systems too small to be considered thermodynamically in a traditional sense  \cite{RubiJPCB2006},  thereby suggesting theoretical work on the rank-size relationships.
  
In future work, it would be of interest to examine whether the UEFA rank  measure  could be used to quantify   a finer description of the  main classes.  Non universality, or class types,  might also be  measured through the  central slope $-2(\psi+\phi).$ In so doing one  might also  imagine weighting performances, see e.g. \cite{JSE10.09.582weightranking}  in  the case of NCAA College Football Rankings,   whence  organizing  more homogeneously   based team competitions, or regulating  various  sport conditions implying  team (but also individual  athletic competitions)   ranking\footnote{Recall horse racing, where an extra weight is put on a horse depending on  previous competitions, in order to level somewhat the field}. This may implies considerations about
  team budgets and expectations, - if more competitiveness is of interest to the organizers.

\bigskip
\begin{flushleft}
{\bf Acknowledgments} 
\end{flushleft}
This work has been performed in the framework of COST Action IS1104 
"The EU in the new economic complex geography: models, tools and policy evaluation".
%

\bigskip

APPENDIX

\bigskip
In this Appendix, in order to illustrate a reviewer question about rank evolution, it is shown that the top teams are not always the same ones. At the end of the 09/10  and  13/14 seasons,  see Table \ref{TableUEFAteammotion}, only 5 teams   (underlined) are both in the top ten ranking. The other teams, either going down or up are mentioned with their   season ranking.

    \begin{table} \begin{center} 
\begin{tabular}[t]{cccccccccc} 
\hline 
Team	&	09/10 rank	&  &Team	& 	13/14 rank	&	\\ \hline	
\underline{FC Barcelona}	&	1	&	down	&	Real Madrid CF	&	1	&	\\	
\underline{Manchester United FC}	&	2	&	down	&	\underline{FC Barcelona}	&	2	&	\\	
\underline{Chelsea FC}	&	3	&	down	&	\underline{FC Bayern Munchen}	&	3	&	\\	
\underline{Arsenal FC}	&	4	&	down	&	\underline{Chelsea FC}	&	4	&	\\	
Liverpool FC	&	5	&	down(*)	&	\underline{Manchester United FC}	&	5	&	\\	
\underline{FC Bayern Munchen}	&	6	&	up	&	 SL Benfica	&	6	&	\\	
Sevilla FC	&	7	&	down(*)	&	Club Atletico de Madrid	&	7	&	\\	
FC Internazionale Milano	&	8	&	down(*)	&	 Valencia CF	&	8	&	\\	
AC Milan	&	9	&	down(*)	&	 \underline{Arsenal FC}	&	9	&	\\	
Olympique Lyonnais	&	10	&	down(*)	&	 FC Porto	&	10	&	\\ \hline	
											
Real Madrid CF	&	13	&	up(**)	&	 AC Milan	&	11	&	\\	
FC Porto	&	15	&	up(**)	&	 Olympique Lyonnais	&	12	&	\\	
SL Benfica	&	17	&	up(**)	&	 FC Internazionale Milano	&	13	&	\\	
Valencia CF	&	20	&	up(**)	&	 Sevilla FC	&	25	&	\\	
Club Atletico de Madrid	&	23	&	up(**)	&	 Liverpool FC	&	32	&	\\ \hline	
\end{tabular} 
   \caption{Illustrating UEFA team rank evolution from 09/10 season to 13/14 season, either going down (*) ,  moving out of the top 10 rank, or going up (**) moving into the top  10 rank   }\label{TableUEFAteammotion}
\end{center} \end{table} 
    \end{document}